\documentclass{article} 
\usepackage{iclr2019_conference}

\usepackage[title]{appendix}

\usepackage{times}
\usepackage{latexsym}

\usepackage{hyperref}

\usepackage{amsmath}
\usepackage{bbm}

\usepackage{amssymb}
\usepackage{algorithm}
\usepackage{algpseudocode}
\usepackage{multirow}
\usepackage{tabu}
\usepackage{booktabs}

\usepackage{graphicx} 
\usepackage{subfigure}
\usepackage{booktabs}       
\usepackage{breakcites}

\usepackage{array}
\newcolumntype{H}{>{\setbox0=\hbox\bgroup}c<{\egroup}@{}}

\iclrfinalcopy

\title{Passage Re-ranking with BERT}

\author{Rodrigo Nogueira\\
New York University\\
\texttt{rodrigonogueira@nyu.edu} \\
\And
Kyunghyun Cho \\
New York University\\
Facebook AI Research\\
CIFAR Azrieli Global Scholar\\
\texttt{kyunghyun.cho@nyu.edu} \\
}

\begin{document}
\maketitle
\begin{abstract}
Recently, neural models pretrained on a language modeling task, such as ELMo~\citep{peters2017semi}, OpenAI GPT~\citep{radford2018improving}, and BERT~\citep{devlin2018bert}, have achieved impressive results on various natural language processing tasks such as question-answering and natural language inference. In this paper, we describe a simple re-implementation of BERT for query-based passage re-ranking. Our system is the state of the art on the TREC-CAR dataset and the top entry in the leaderboard of the MS MARCO passage retrieval task, outperforming the previous state of the art by 27\% (relative) in MRR@10. The code to reproduce our results is available at \url{https://github.com/nyu-dl/dl4marco-bert}
\end{abstract}

\section{Introduction}

We have seen rapid progress in machine reading compression in recent years with the introduction of large-scale datasets, such as SQuAD~\citep{rajpurkar2016squad}, MS MARCO~\citep{nguyen2016ms}, SearchQA~\citep{dunn2017searchqa}, TriviaQA~\citep{joshi2017triviaqa}, and QUASAR-T~\citep{dhingra2017quasar}, and the broad adoption of neural models, such as BiDAF~\citep{seo2016bidirectional}, DrQA~\citep{chen2017reading}, DocumentQA~\citep{clark2017simple}, and QAnet~\citep{yu2018qanet}.

The information retrieval (IR) community has also experienced a flourishing development of neural ranking models, such as DRMM~\citep{guo2016deep}, KNRM~\citep{xiong2017end}, Co-PACRR~\citep{hui2018co}, and DUET~\citep{mitra2017learning}. However, until recently, there were only a few large datasets for passage ranking, with the notable exception of the TREC-CAR~\citep{dietz2017trec}. This, at least in part, prevented the neural ranking models from being successful when compared to more classical IR techniques~\citep{lin2019neural}.

We argue that the same two ingredients that made possible much progress on the reading comprehension task are now available for passage ranking task. Namely, the MS MARCO passage ranking dataset, which contains one million queries from real users and their respective relevant passages annotated by humans, and BERT, a powerful general purpose natural language processing model.

In this paper, we describe in detail how we have re-purposed BERT as a passage re-ranker and achieved state-of-the-art results on the MS MARCO passage re-ranking task.

\section{Passage Re-Ranking with BERT}

\paragraph{Task}

A simple question-answering pipeline consists of three main stages. First, a large number (for example, a thousand) of possibly relevant documents to a given question are retrieved from a corpus by a standard mechanism, such as BM25. In the second stage, \textit{passage re-ranking}, each of these documents is scored and re-ranked by a more computationally-intensive method. Finally, the top ten or fifty of these documents will be the source for the candidate answers by an answer generation module. 
In this paper, we describe how we implemented the second stage of this pipeline, passage re-ranking.

\paragraph{Method}

The job of the re-ranker is to estimate a score $s_i$ of how relevant a candidate passage $d_i$ is to a query $q$.
We use BERT as our re-ranker. Using the same notation used by~\citet{devlin2018bert}, we feed the query as sentence A and the passage text as sentence B. We truncate the query to have at most 64 tokens. We also truncate the passage text such that the concatenation of query, passage, and separator tokens have the maximum length of 512 tokens. We use a $\text{BERT}_\text{LARGE}$ model as a binary classification model, that is, we use the $[\text{CLS}]$ vector as input to a single layer neural network to obtain the probability of the passage being relevant. We compute this probability for each passage independently and obtain the final list of passages by ranking them with respect to these probabilities. 

We start training from a pre-trained BERT model and fine-tune it to our re-ranking task using the cross-entropy loss: 
\begin{equation} 
\label{eq:aggregator_loss}
L = -\sum_{j \in J_{\text{pos}}} \log (s_j) - \sum_{j \in J_{\text{neg}}} \log (1 - s_j),
\end{equation}
where $J_{\text{pos}}$ is the set of indexes of the relevant passages and $J_{\text{neg}}$ is the set of indexes of non-relevant passages in top-1,000 documents retrieved with BM25. 

\section{Experiments}

We train and evaluate our models on two passage-ranking datasets, MS MARCO and TREC-CAR.

\subsection{MS MARCO}

The training set contains approximately 400M tuples of a query, relevant and non-relevant passages. The development set contains approximately 6,900 queries, each paired with the top 1,000 passages retrieved with BM25 from the MS MARCO corpus. On average, each query has one relevant passage. However, some have no relevant passage because the corpus was initially constructed by retrieving the top-10 passages from the Bing search engine and then annotated. Hence, some of the relevant passages might not be retrieved by BM25.

An evaluation set with approximately 6,800 queries and their top 1,000 retrieved passages without relevance annotations is also provided.

\paragraph{Training}

We fine-tune the model using a TPU v3-8\footnote{
\url{https://cloud.google.com/tpu/}
} 
with a batch size of 128 (128 sequences * 512 tokens = 65,536 tokens/batch) for 100k iterations, which takes approximately 30 hours. This corresponds to training on 12.8M (100k * 128) query-passage pairs or \textit{less than 2\% of the full training set}. We could not see any improvement in the dev set when training for another 3 days, which equivalent to seeing 50M pairs in total.

We use ADAM~\citep{kingma2014adam} with the initial learning rate set to $3 \times 10^{-6}$, $\beta_1 = 0.9$, $\beta_2 = 0.999$, L2 weight decay of 0.01, learning rate warmup over the first 10,000 steps, and linear decay of the learning rate. We use a dropout probability of $0.1$ on all layers.

\begin{table*}[t]
\begin{center}
\begin{tabular}{l|cc|c}
 & \multicolumn{2}{c|}{MS MARCO} & TREC-CAR\\
 & \multicolumn{2}{c|}{MRR@10} & MAP\\
Method & Dev & Eval & Test\\
\noalign{\vskip 1mm}
\toprule
\noalign{\vskip 1mm}
BM25 (Lucene, no tuning) & 16.7 & 16.5 & 12.3 \\
BM25 (Anserini, tuned) & - & - & 15.3 \\
Co-PACRR$^\star$~\citep{macavaneycontextualized} & - & - & 14.8\\
KNRM~\citep{xiong2017end} & 21.8 & 19.8 & -\\
Conv-KNRM~\citep{dai2018convolutional} & 29.0 & 27.1 & -\\
IRNet$^\dagger$ & 27.8 &  28.1 & -\\
\noalign{\vskip 1mm}
\midrule
\noalign{\vskip 1mm}
BERT Base & 34.7 & - & 31.0\\
BERT Large & 36.5 & 35.8 & 33.5\\
\end{tabular}
\end{center}
\vskip -1mm
\caption{Main Result on the passage re-ranking datasets. \mbox{$\star$ Best Entry in the TREC-CAR 2017}. \mbox{$\dagger$ Previous SOTA in the MS MARCO leaderboard as of 01/04/2019}; unpublished work.}
\label{tab:results_document_retrieval}
\end{table*}

\subsection{TREC-CAR}

Introduced by~\citet{dietz2017trec}, in this dataset, the input query is the concatenation of a Wikipedia article title with the title of one of its section. The relevant passages are the paragraphs within that section. The corpus consists of all of the English Wikipedia paragraphs, except the abstracts. The released dataset has five predefined folds, and we use the first four as a training set (approximately 2.3M queries), and the remaining as a validation set (approximately 580k queries). The test set is the same one used to evaluate the submissions to TREC-CAR 2017 (approx. 2,254 queries).

Although TREC-CAR 2017 organizers provide manual annotations for the test set, only the top five passages retrieved by the systems submitted to the competition have manual annotations. This means that true relevant passages are not annotated if they rank low. Hence, we evaluate using the automatic annotations, which provide relevance scores for all possible query-passage pairs.

\paragraph{Training}

We follow the same procedure described for the MS MARCO dataset to fine-tune our models on TREC-CAR. However, there is an important difference. The official pre-trained BERT models\footnote{
\url{https://github.com/google-research/bert}
} 
were pre-trained on the full Wikipedia, and therefore they have seen, although in an unsupervised way, Wikipedia documents that are used in the test set of TREC-CAR. Thus, to avoid this leak of test data into training, we pre-trained the BERT re-ranker only on the half of Wikipedia used by TREC-CAR's training set.

For the fine-tuning data, we generate our query-passage pairs by retrieving the top ten passages from the entire TREC-CAR corpus using BM25.\footnote{We use the Anserini toolkit~\citep{yang2018anserini} to index and retrieve the passages.} This means that we end up with 30M example pairs (3M queries * 10 passages/query) to train our model. We train it for 400k iterations, or 12.8M examples (400k iterations * 32 pairs/batch), which corresponds to only 40\% of the training set. Similarly to MS MARCO experiments, we did not see any gain on the dev set by training the models longer.

\subsection{Results}

We show the main result in
Table~\ref{tab:results_document_retrieval}. Despite training on a fraction of the data available, the proposed BERT-based models surpass the previous state-of-the-art models by a large margin on both of the tasks.

\begin{figure}
\begin{center}
\centerline{\includegraphics[width=0.7\textwidth]{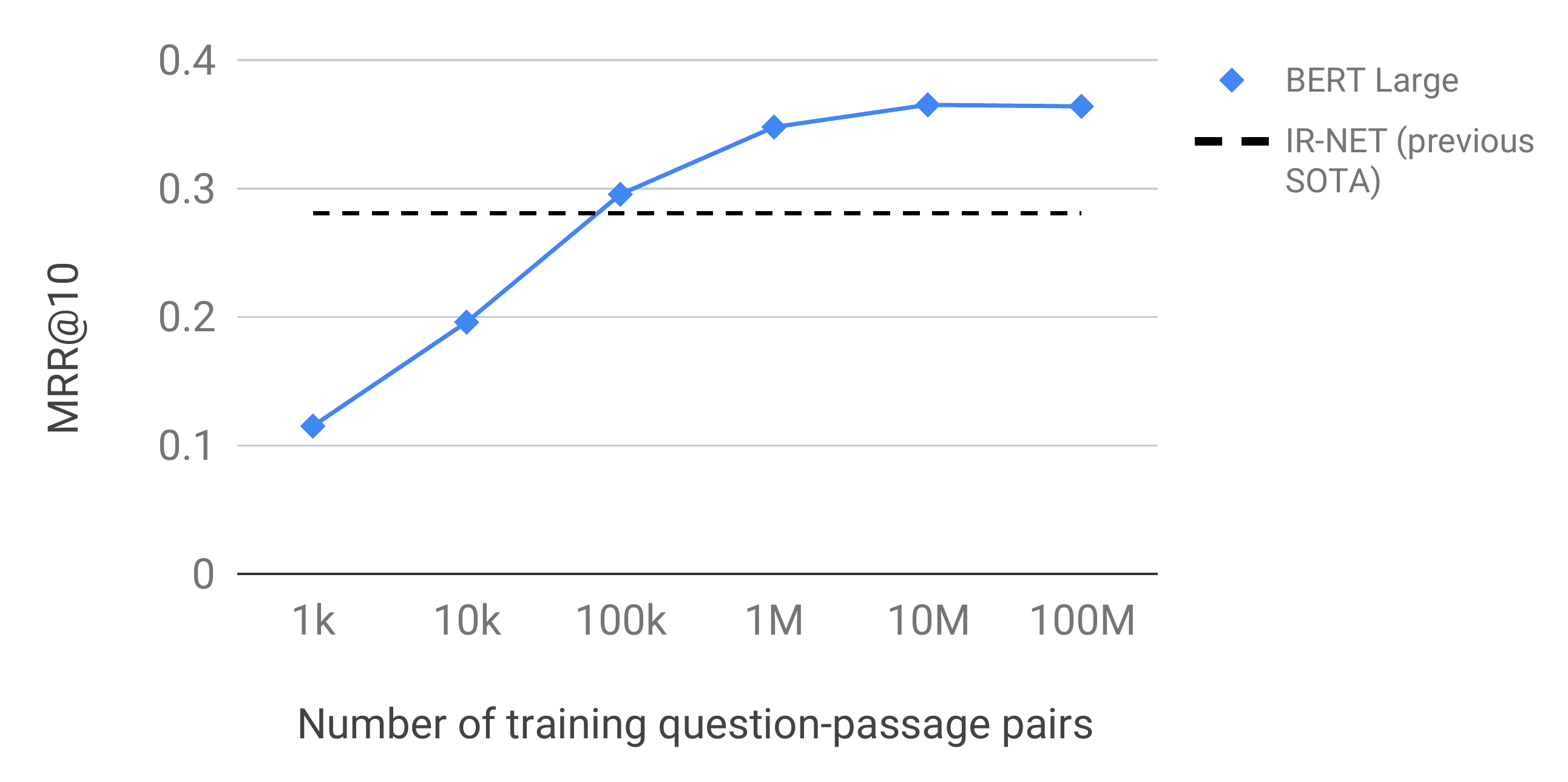}}
\vspace{-4mm}
\caption{Number of MS MARCO examples seen during training vs. MRR@10 performance.} 
\label{fig:training_size}
\end{center}
\vspace{-6mm}
\end{figure}

\paragraph{Training size vs performance:} We found that the pretrained models used in this work require few training examples from the end task to achieve a good performance~\ref{fig:training_size}. For example, a $\text{BERT}_\text{LARGE}$ trained on 100k question-passage pairs (less than 0.3\% of the MS MARCO training data) is already 1.4 MRR@10 points better than the previous state-of-the-art, IR-NET.

\section{Conclusion}

We have described a simple adaptation of BERT as a passage re-ranker that has become the state of the art on two different tasks, which are TREC-CAR and MS MARCO. We have made the code to reproduce our experiments publicly available.

\bibliographystyle{iclr2019_conference}
\bibliography{main}

\end{document}